\documentclass[preprint,showpacs,preprintnumbers,amsmath,amssymb]{revtex4}
\usepackage{graphics,epsfig,subfigure}
\usepackage{color}

\begin{document}
\renewcommand{\baselinestretch}{1.3}
\newcommand\beq{\begin{equation}}
\newcommand\eeq{\end{equation}}
\newcommand\beqn{\begin{eqnarray}}
\newcommand\eeqn{\end{eqnarray}}
\newcommand\nn{\nonumber}
\newcommand\fc{\frac}
\newcommand\lt{\left}
\newcommand\rt{\right}
\newcommand\pt{\partial}

\allowdisplaybreaks

\title{Constructing higher dimensional exact black holes in Einstein-Maxwell-scalar-theory}

\author{Jianhui Qiu\footnote{jhqiu@nao.cas.cn}\ \ \ \ Changjun Gao\footnote{gaocj@bao.ac.cn}}
\affiliation{
Key Laboratory of Computational Astrophysics, National Astronomical Observatories,
Chinese Academy of Sciences, Beijing 100101, China and
School of Astronomy and Space Sciences, University of Chinese Academy of Sciences,
No. 19A, Yuquan Road, Beijing 100049, China}

\begin{abstract}
We construct higher dimensional and exact black holes in Einstein-Maxwell-scalar-theory. The strategy we adopted is to extend the known, static and spherically symmetric black holes in the Einstein-Maxwell-dilaton gravity and Einstein-Maxwell-scalar theory. Then we investigate the black hole thermodynamics. Concretely, the generalized Smarr formula and the first law of thermodynamics are derived.

\end{abstract}








\maketitle



\section{Introduction}
According to the Lovelock \cite{lovelock1971einstein,lovelock1972four} and the Ostrogradsky instability \cite{woodard2015theorem} theorems, it is uniquely the Einstein gravity  that consists of metric and its derivatives and with the equations of motion no more than second order. So in order to extend the Einstein gravity, the easiest way is to couple it with a scalar field. On the other hand, string theory is generally considered to be the most promising approach to unify quantum theory and gravity in higher dimensions. The low energy limit of string theory does lead to the Einstein gravity coupled non-minimally to a scalar dilaton field \cite{green1987superstring}. The dilaton field, coupled in a nontrivial way to other fields such as gauge fields has aroused many interests and many black hole solutions are found \cite{gibbons1988black,garfinkle1991charged,brill1991negative,gregory1993black}.

These solutions are all asymptotically flat. It has been proved \cite{poletti1994global,poletti1995charged} that, in the presence of one or two Liouville-type potential which is considered as a generalization of the cosmological constant, neither asymptotically flat nor (anti)-de Sitter solutions exist. However, by combining three Liouville type dilaton potential, one successfully constructs a higher dimensional asymptotically (anti)-de Sitter solutions \cite{gao2005higher}. Then the topological anti-de Sitter black branes  with higher dimensions in Einstein-Maxwell-dilaton theory were constructed and their properties were investigated \cite{hendi2010thermodynamics}. With the same dilaton potential in \cite{gao2005higher},  Sheykhi \cite{sheykhi2010higher} finds the metric for the n-dimensional charged slowly rotating dilaton black hole in the background of asymptotically (anti)-de Sitter spacetime.

On the other hand, a remarkable phenomenon of spontaneous scalarisation of charged black holes is recently discovered \cite{herdeiro2018spontaneous,fernandes2019spontaneous} and vast studies on the scalarisation of black holes in various Einstein-Maxwell-scalar (EMS) models (see \cite{blazquez2020einstein} and references therein) are carried out. Most of these studies are based on numerical calculations. In view of this point, we will construct exact charged black hole solutions in EMS theory. Our strategy is to extend the known, static and spherically symmetric black holes in the Einstein-Maxwell-dilaton gravity to EMS theory. Actually, using this method, we have constructed the four dimensional black holes in EMS in \cite{yu2020constructing}. Thus the purpose of this paper is to extend it from four dimensions to higher dimensions.

The paper is organized as follows. In Sec.~II, we derive the equations of motion for the fields and present the metric. In Sec.~III, we introduce the generalized Smarr formula for this solution and verify the validity of the first law of black hole thermodynamics. In Sec.~IV, we investigate the thermodynamic stability problem and the phase transitions by using heat capacity and Gibbs free energy. In Sec.~IV we summarize our results.

 \section{ACTION AND THE EQUATIONS OF MOTION}

We start from the action of Einstein-Maxwell-scalar theory
\beq
S=\int{}d^nx\sqrt{-g}\left[R-\frac{4}{n-2}\nabla_\mu\phi\nabla^\mu\phi-V(\phi)-K(\phi)F^2\right]\;,
\label{action}
\eeq
where $R$ is the Ricci scalar curvature, $F^2\equiv F_{\mu\nu}F^{\mu\nu}$ comes from the
Maxwell field,  $K\left(\phi\right)$ is the coupling function between scalar field and Maxwell field.  $V\left(\phi\right)$ is the scalar potential.

Varying the action with respect to the metric, Maxwell and the scalar field, respectively, yields

\beq\label{g0}
R_{\mu\nu}=\frac{4}{n-2}\nabla_\mu\phi\nabla_\nu\phi+\frac{V}{n-2}g_{\mu\nu}+2KF_{\mu\alpha}F_\nu^\alpha-\frac{K}{n-2}F^2g_{\mu\nu})\;,
\eeq

\beq\label{F0}
\partial_\mu\left(\sqrt{-g}KF^{\mu\nu}\right)=0\;,
\eeq

\beq\label{p0}
\nabla_\mu\nabla^\mu\phi-\frac{n-2}{8}\left( \frac{\partial V}{\partial\phi}+\frac{\partial K}{\partial\phi}F^2\right)=0\;.
\eeq

We choose the most general form of the metric for static black hole as follows
\beq\label{metric0}
ds^2=-U\left(x\right)dt^2+\frac{1}{U\left(x\right)}dx^2+f\left(x\right)^2d\Omega_{n-2}^{2}\;,
\eeq
where $x$ denotes the radial variable. Then the Maxwell equation (\ref{F0}) can be integrated to give
\beq\label{maxwell0}
F^{10}=\frac{q}{Kf^{n-2}}\;,
\eeq
where $q$ is the constant of integration and it has the dimension of $l^{n-3}$.  Then the equations of motion (\ref{g0}) to (\ref{p0}) reduce to three independent equations
\beq
\label{g1}
\frac{1}{f^{n-2}}\frac{d}{dx}\left(f^{n-2}U\frac{d\phi}{dx}\right)=\frac{n-2}{8}\left[\frac{\partial V}{\partial\phi}-\frac{2q^2\partial_\phi K}{f^{2n-4}K^2}\right]\;,
\eeq
\beq
\label{F1}
\frac{1}{f}\frac{d^2f}{dx^2}=-\frac{4}{\left(n-2\right)^2}\left(\frac{d\phi}{dx}\right)^2\;,
\eeq
\beq
\label{p1}
\frac{1}{f^{n-2}}\frac{d}{dx}\left(U\frac{d}{dx}f^{n-2}\right)=\frac{\left(n-2\right)\left(n-3\right)}{f^2}-V-\frac{2q^2}{Kf^{2n-4}}\;.
\eeq
There are five functions $U$, $f$, $\phi$, $V$ and $K$ in these equations. But we have only three equations. Therefore, the system of equations are not closed. In general, one usually presumes $V,\ K$ and then solve for $U,\ f,\ \phi$. For example, when we assume \cite{gao2005higher}
\begin{eqnarray}
K\left(\phi\right)&=&e^{\frac{4\alpha\phi}{n-2}}\;,
\end{eqnarray}
\begin{eqnarray}\label{VV}
V\left(\phi\right)&=&\frac{\lambda}{3\left(n-3+\alpha^2\right)^2}
\left[-\alpha^2\left(n-2\right)\left(n^2-n\alpha^2-6n+\alpha^2+9\right)
e^{\frac{4\left(n-3\right)\phi}{\left(n-2\right)\alpha}}\right.\nonumber
\\ &&\left. +\left(n-2\right)
\left(n-3\right)^2\left(n-1-\alpha^2\right)e^{-\frac{4\alpha\phi}{n-2}}+4\alpha^2\left(n-3\right)
\left(n-2\right)^2e^{\frac{2\phi\left(n-3-\alpha^2\right)}{\left(n-2\right)\alpha}}\right]\;,
\end{eqnarray}
we will have the solution as follows \cite{gao2005higher}
\begin{eqnarray}
\label{ndd}
d{s}^2&=&-\left\{\left[1-\left(\frac{b}{{r}}\right)^{n-3}\right]
\left[1-\left(\frac{a}{{r}}\right)^{n-3}\right]^{1-\gamma\left(n-3\right)}
-\frac{1}{3}\lambda
r^2\left[1-\left(\frac{a}{{r}}\right)^{n-3} \right]^{\gamma}
\right\}d{t}^2\nonumber\\&&+\left\{\left[1-\left(\frac{b}{{r}}\right)^{n-3}\right]
\left[1-\left(\frac{a}{{r}}\right)^{n-3}\right]^{1-\gamma\left(n-3\right)}
-\frac{1}{3}\lambda
r^2\left[1-\left(\frac{a}{{r}}\right)^{n-3} \right]^{\gamma}
\right\}^{-1}\nonumber
\\ &&\cdot\left[1-\left(\frac{a}{{r}}\right)^{n-3}
\right]^{-\gamma\left(n-4\right)}d{r}^2+
r^2\left[1-\left(\frac{a}{{r}}\right)^{n-3}\right]^{\gamma}d\Omega_{n-2}^2\;,
\end{eqnarray}
\begin{equation}
\phi=-\frac{(n-2) \alpha}{2\left(\alpha^{2}+n-3\right)} \ln \left[1-\left(\frac{a}{r}\right)^{n-3}\right]\;.
\end{equation}
Here $a,\ b$ are two integration constants which are related to the black hole mass and electric charge. $\lambda$ is the cosmological constant and $\alpha$ is the coupling constant. $\gamma$ is given by \cite{gao2005higher}
\begin{eqnarray}
\gamma &\equiv &\frac{2\alpha^2}{\left(n-3\right)\left(n-3+\alpha^2\right)}\;.
\end{eqnarray}
Different from above example, here we shall presume $U,\ f$ in advance and then solve for $\phi,\ V,\ K$. In order to find the desirable expressions for $U$ and $f$, we observe Eq.~(\ref{ndd}) and find that the $\lambda$ term is proportional to $r^2\left[1-\left(\frac{a}{{r}}\right)^{n-3}\right]^{\gamma}$.

On the other hand, the four dimensional black hole solution in EMS theory is \cite{yu2020constructing}

\begin{equation}\label{eq:u2a}
U=\left(1-\frac{2m}{r}\right)\left(1-\frac{Q^2}{mr}\right)^{\frac{1-\alpha^2}{1+\alpha^2}}+\frac{\beta Q^2}{f^2}-\frac{1}{3}\lambda f^2\;,
\end{equation}
\begin{equation}\label{eq:u2b}
f=r\left(1-\frac{Q^2}{mr}\right)^{\frac{\alpha^2}{1+\alpha^2}}\;,
\end{equation}
\begin{equation}\label{eq:2d}
K=\frac{2e^{2\alpha\phi}}{2+\beta+\beta \alpha^2 e^{2\alpha\phi+2\phi/\alpha}}\;,
\end{equation}
\begin{eqnarray}\label{eq:2a}
V&=&\frac{2\lambda}{3\left(1+\alpha^2\right)^2}\left[\alpha^2\left(3\alpha^2-1\right)e^{2\phi/\alpha}+\left(3-\alpha^2\right)e^{-2\alpha\phi}\right.\nonumber\\&&\left.+8\alpha^2 e^{-\phi\alpha+\phi/\alpha}\right]\;.
\end{eqnarray}
Therefore, motivated by Eq.~(\ref{ndd}) and Eq.~(\ref{eq:u2a}), we presume a new solution

\begin{eqnarray}
\label{ns}
d{s}^2&=&-Ud{t}^2+\frac{1}{U}\cdot\left[1-\left(\frac{a}{{r}}\right)^{n-3}
\right]^{-\gamma\left(n-4\right)}d{r}^2+
f^2d\Omega_{n-2}^2\;,
\end{eqnarray}
with
\begin{eqnarray}\label{uf}
U&=&\left[1-\left(\frac{b}{{r}}\right)^{n-3}\right]
\left[1-\left(\frac{a}{{r}}\right)^{n-3}\right]^{1-\gamma\left(n-3\right)}
-\frac{1}{3}\lambda
r^2\left[1-\left(\frac{a}{{r}}\right)^{n-3} \right]^{\gamma}\nonumber\\&&+{\beta\,{q}^{2} \left\{r^{2}
\left[ 1- \left( {\frac {a}{r}} \right) ^{n-3} \right]^{\gamma}
\right\} ^{3-n}}\;,\nonumber\\
f&=&r\left[1-\left(\frac{a}{{r}}\right)^{n-3}\right]^{\gamma/2}\;,
\end{eqnarray}
where $\beta$ is a constant. Now a new term of $\beta$ is inserted in the expression of $U$. Since $q$ has the dimension of $l^{n-3}$, $\beta$ is dimensionless.  Given the expressions of $U,\ f$, the expressions of $\phi,\ K,\ V$ are then worked out from the equations of motion. To this end, we transform the equations of motion from $(t,\ x)$ coordinates to $(t,\ r)$ coordinates
via the following coordinates transformation
\beq
\label{cotr}
x=\int{} dr\left[1-(\frac{r_{-}}{r})^{n-3} \right]^{-\gamma (n-4)/2}\;,\ \ \ \textrm{or} \ \ \  r'=\left[1-\left(\frac{r_{-}}{r}\right)^{n-4} \right]^{\gamma (n-4)/2}\;.
\eeq
Then the equations of motion Eqs.~(\ref{g1},\ \ref{F1},\ \ref{p1}) turn out to be

\beq
\label{g2}
\frac{1}{f^{n-2}}r'\frac{d}{dr}\left(f^{n-2}Ur'\frac{d\phi}{dr}\right)=\frac{n-2}{8}\left(\frac{\partial V}{\partial\phi}-\frac{2q^2\partial_\phi K}{f^{2n-4}K^2}\right)\;,
\eeq
\beq
\label{F2}
\frac{1}{f}\frac{d}{dr}(r'\frac{df}{dr})=-\frac{4}{(n-2)^2}\left(\frac{d\phi}{dr}\right)^2r'\;,
\eeq
\beq
\label{p2}
\frac{1}{f^{n-2}}r'\frac{d}{dr}\left(Ur'\frac{d}{dr}f^{n-2}\right)=\frac{(n-2)(n-3)}{f^2}-V-\frac{2q^2}{Kf^{2n-4}}\;.
\eeq
Substituting Eqs.~(\ref{uf}) into above equations of motion, we obtain

\begin{equation}
\phi=-\frac{(n-2) \alpha}{2\left(\alpha^{2}+n-3\right)} \ln \left[1-\left(\frac{a}{r}\right)^{n-3}\right]\;,
\end{equation}

\beq
\begin{split}
K(\phi)=&\frac{2\left( {\alpha}^{2}+n-3 \right){{\rm e}^{{\frac {4\alpha\phi}{n-2}}}}}{ 2\,{\alpha}^{2}+ \left( n-3 \right)  \left[ 2+ \left( {n}^{2}
			-5\,n+6 \right) \beta \right] +\beta\,{
		\alpha}^{2} \left( n-2 \right)  \left(n-3 \right){{\rm e}^{{\frac { 4\left(
			{\alpha}^{2}+n-3 \right)\phi }{\alpha\, \left( n-2 \right) }}}}}\;,
\end{split}
\eeq

and
\begin{equation}
\begin{array}{l}
q^{2}=\frac{(n-2)(n-3)^{2}}{2\left(n-3+\alpha^{2}\right)} a^{n-3} b^{n-3}\;,
\end{array}
\end{equation}
with $V$ the exact form of Eq.~(\ref{VV}). When $n=4$, $K$ restores to Eq.~(\ref{eq:2d}). Up to this point, the n-dimensional and exact
black hole solution is constructed in EMS theory.

In the next, we calculate the electric charge and mass of the black hole. The electric charge is
\beq
Q=\frac{1}{4\pi}\lim_{x \to \infty}\int{} F_{tr}\sqrt{-g} K\left(\phi\right)d^{n-2}x=\frac{\Omega_{n-2}}{4\pi}q\;.
\eeq
In order to calculate the conserved mass of the spacetime, we use the subtraction method of Brown and York \cite{brown1993quasilocal,brown1994temperature}. The definition of conserved mass is given by
\begin{equation}
{M}\equiv \frac{1}{8 \pi} \int_{\mathcal{B}} d^{2} \varphi \sqrt{\sigma}\left\{\left(K_{a b}-K h_{a b}\right)-\left(K_{a b}^{0}-K^{0} h_{a b}^{0}\right)\right\} n^{a} \xi^{b}\;,
\end{equation}
where $\mathcal{B}$ is the boundary surface of the spacetime, $\xi$ a timelike Killing vector field on $\mathcal{B}$, $\sigma$ the determinant of the metric of the boundary $\mathcal{B}$, $K_{a b}^{0}$ the extrinsic curvature of the background metric and $n^{a}$ the timelike unit normal vector to the boundary $\mathcal{B}$. In  the context of counterterm method and following the procedure of \cite{sheykhi2010thermodynamic}, we get, after a detailed calculation,
\beq
M=\frac{\Omega_{n-2}}{16\pi}\left(n-2\right)\left[b^{n-3}+\left(1-\left(n-3\right)\gamma\right)a^{n-3}\right]\;,
\eeq
which is the same as \cite{sheykhi2010thermodynamic}.

 \section{THERMODYNAMICS}

In this section, we explore the black hole thermodynamics. Concretely, we shall construct the generalized Smarr formula and the first law of thermodynamics. To this end, we start from the calculation of Hawking temperature which is defined as follows:
 \beq
\kappa^2=-\frac{1}{2}\nabla_a\chi_b\nabla^a\chi^b\;,
\eeq
where $\chi^\mu$ is a Killing vector field which is null on the horizon. Since we are dealing with a static metric and we can chose $\chi^\mu=\frac{\partial}{\partial t}$.  Then one can write the expression of temperature for black holes
\beq
T=\frac{1}{4\pi}\frac{U'(r_{+})}{\sqrt{U(r_{+})W(r_{+})}}\;,
\eeq
with the metric
\begin{equation}
\begin{aligned}
ds^{2}=-U(r)dt^2+W(r)dr^{2}+f(r)^2d\Omega_{n-2}^{2}\;.
\end{aligned}
\end{equation}
After substituting the equation of event horizon $U(r_{+})=0$
, we get the formula for temperature

\begin{equation}
\begin{aligned}
T=& \frac{1}{12 \pi r_+}\left[-3 q^{2} \beta r_{+}^{-2 n+6}(n-3) \Gamma^{-1+\frac{(-n+2)\gamma}{2}}+(-9+3 n) \Gamma^{1+\frac{(-n+2)\gamma}{2}}\right.\\
&\left.+r_{+}^{2} \lambda\left(-(n-3)\left(\gamma n-2\gamma-1\right)\right)\Gamma^{-1+\frac{\gamma(-2+n)}{2}}+\Gamma^{\frac{\gamma(n-2)}{2}}(n-2)\left(\gamma n-3 \gamma-2\right)\right]\;,
\end{aligned}
\end{equation}
where $\Gamma \equiv 1-\left( \frac{a}{r_{+}} \right)^{n-3}$ and $r_{+}$ represents the radius of black hole event horizon.

The entropy of black holes generally satisfies the area law which states that the entropy is a quarter of the area of black hole event horizon \cite{bekenstein1973black,hawking1974nature,gibbons1977cosmological}. This nearly universal law applies to almost all kinds of black holes, including Einstein-Maxwell-scalar black  holes \cite{hunter1998action,hawking1999nut,mann1999misner}. Therefore, we have the entropy

\beq
\begin{split}
S=\frac{A}{4}=\frac{f(r_{+})^{n-2}\Omega_{n-2}}{4}\;.
\end{split}
\eeq

The electrical potential is defined by
\begin{equation}
\begin{aligned}
\Phi&=\int_{x_{1}}^{\infty} \frac{d A_{0}}{d x} d x=\int_{r_{+}}^{\infty} \frac{d A_{0}}{d x} \frac{d x}{d r} d r=\int_{r_{+}}^{\infty} -F^{10}\left[1-\left(\frac{a}{r}\right)^{n-3}\right]^{-\frac{\gamma(n-4)}{2}}dr\\& =\int_{r_{+}}^{\infty} \frac{-q}{f(r)^{n-2}K(\phi)}\left[1-\left(\frac{a}{r}\right)^{n-3}\right]^{-\frac{\gamma(n-4)}{2}}dr\\&=\eta/\left[2(-3+n)\left(\alpha^{2}+n-3\right)\left(a^{3} r^{n}-a^{n} r_{+}^{3}\right)\right]\;,
\end{aligned}
\end{equation}
where $\eta$ is defined by
\begin{equation}
\begin{aligned}
&\eta \equiv q\left[-a^{n}\left(n^{3} \beta-8n^{2} \beta+(21 \beta+2) n+2 \alpha^{2}-18 \alpha-6\right) r_+^{6-n}\right.\\ &\left. \qquad +a^{3} r_+^{3}\left(\alpha^{2}+n-3\right)\left(n^{2} \beta-5 \beta n+6 \beta+2\right)\right]\;.
\end{aligned}
\end{equation}
When $\beta=0$, $\Phi$ reduces to that in \cite{sheykhi2010thermodynamic}.

We define the thermal pressure $P$
\begin{equation}
P=\frac{-1}{8\pi}\frac{(n-1)(n-2)\lambda}{6}\;.
\end{equation}
Given the equation of event horizon, $\lambda $ can be expressed as the functions of $a,\ b,\ r_{+}$. But we don't bother to give it here because it is too lengthy.

The conjugate thermal volume is
\begin{equation}
\begin{split}
\mathfrak{V}&\equiv\left(\frac{\partial M}{\partial P}\right)_{S, Q}=\left( n-2 \right) {r_{+}}^{-1+n} \left( 1-{\frac {a}{r_{+}}} \right) ^{{
		\frac {2{\alpha}^{2} \left( n-2 \right) }{ \left( -3+n \right)  \left(
			{\alpha}^{2}+n-3 \right) }}}\\ & \cdot {\frac {1}{6\left({\alpha}^{2}+n-3\right)} \left[ -3+n+{\alpha}^{2}- \left( -3+
		n \right)  \left( {\frac {a}{r_{+}}} \right) ^{-3+n} \right]  \left[ 1-
	\left( {\frac {a}{r_{+}}} \right) ^{-3+n} \right] ^{-1}}\;.
\end{split}
\end{equation}
Then we find that the generalized Smarr formula
\begin{equation}
(n-3) M=(n-2) T S+(n-3) \Phi Q-2 P \mathfrak{V}\;,
\end{equation}
is indeed satisfied. It is apparent the formula is related to dimension of spacetime.

Choosing $a$, $b$, $r_{+}$ as independent variables and making the differentiation $M$, $S$, $Q$, $ P$ with respect to $a$, $b$, $r_{+}$, we obtain
\begin{equation}
\begin{array}{l}
d M=\frac{\partial M}{\partial a} d a+\frac{\partial M}{\partial b} d b+\frac{\partial M}{\partial r_{+}} d r_{+}\;, \\
d S=\frac{\partial S}{\partial a} d a+\frac{\partial S}{\partial b} d b+\frac{\partial S}{\partial r_{+}} d r_{+}\;, \\
d Q=\frac{\partial Q}{\partial a} d a+\frac{\partial Q}{\partial b} d b+\frac{\partial Q}{\partial r_{i}} d r_+\;, \\
d P=\frac{\partial P}{\partial a} d a+\frac{\partial P}{\partial b} d b+\frac{\partial P}{\partial r_{+}} d r_{+}\;.
\end{array}
\end{equation}
After straightforward but complicated calculations, the first law of thermodynamics $dU=TdS+\Phi dQ-\mathfrak{V}dP$ is indeed satisfied. Different from the Smarr formula, the first law is not related to the dimension of spacetime.

\section{HEAT CAPACITY AND STABILITY IN CANONICAL ASSEMBLE}
The local stability of a thermodynamical system in canonical ensemble depends on the sign of the heat capacity. If the sign is positive, the system is thermodynamically stable. On the contrary, if the sign is negative,  the system would go under phase transition and then acquires stable state. This phase transition could happen whenever heat capacity meets a root or has a divergency.

\begin{figure}[htbp]
\label{alpha2CQT}
\centering
\subfigure[]{\label{alpha2CQ-r}
\includegraphics[width=8cm,height=6cm]{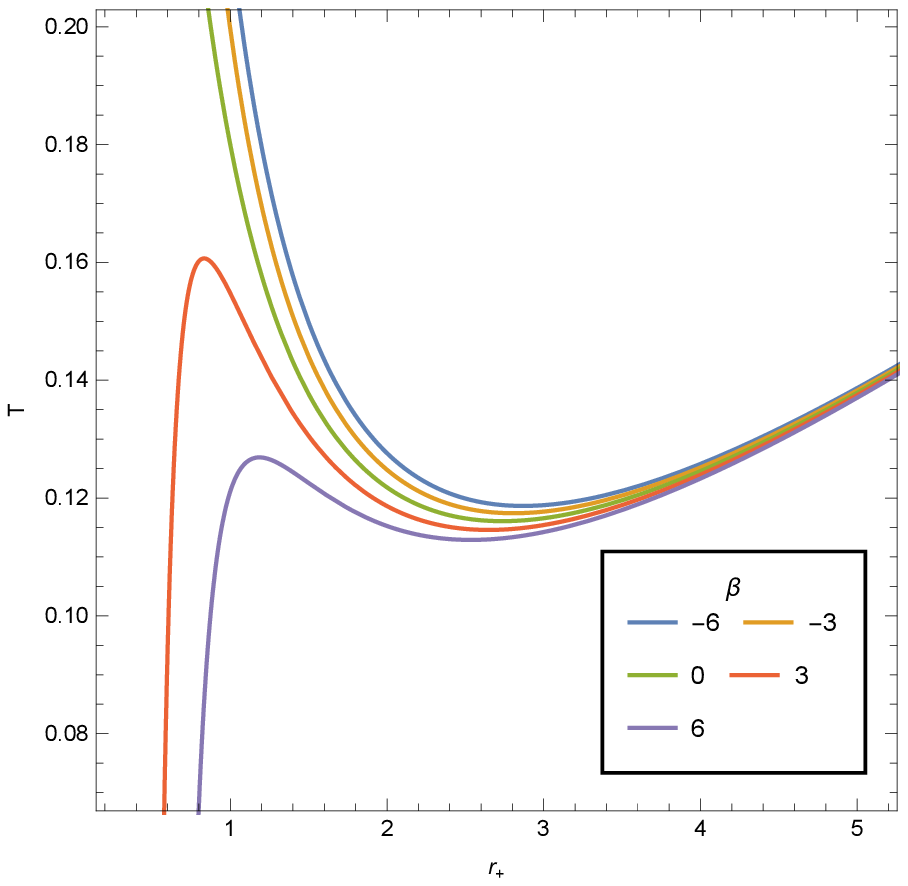}}
\subfigure[ ]{  \label{alpha2T-r}
\includegraphics[width=8cm,height=6cm]{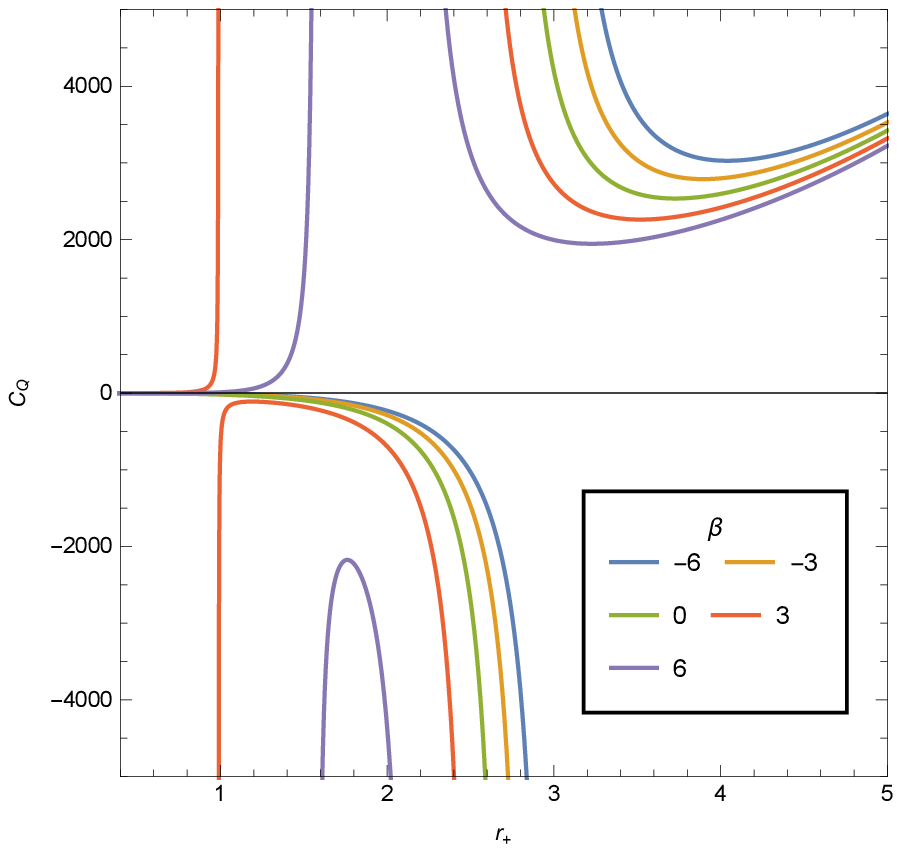}}\\
\caption{(a) The black hole heat capacity with respect to the event horizon $r_{+}$. (b) The black hole temperature with respect to event horizon $r_{+}$.  The parameters are $n=5$, $\alpha=2$, $a=0.2$, $\lambda=-0.2$.}
\end{figure}

In Fig.~\ref{alpha2CQ-r} and Fig.~\ref{alpha2T-r}, we plot the black hole temperature $T$ and the heat capacity $C_Q$ with respect to the black hole event horizon $r_+$. We consider the parameters as follows $n=5,\ \alpha=2,\ a=0.2,\ \lambda=-0.2$ for different $\beta$. The negative $\lambda$ means we are considering black hole in anti-Sitter universe. Fig.~\ref{alpha2CQ-r} shows that when $\beta\leq 0$, there are two phases of black holes, the so-called small black holes and large black holes, respectively. On the other hand, if $\beta>0$, there would be three phases of black holes, the so-called small, middle and large black holes, respectively. The case with positive $\beta$ has generally two event horizons, namely, the inner horizon and the outer event horizon while the negative $\beta$ has only one outer event horizon. Fig.~\ref{alpha2T-r} shows some points of divergence. According to the viewpoint of Davies \cite{davies1977thermodynamic}, the divergence of heat capacity means the second-order phase transition. Comparing it with Fig.~\ref{alpha2CQ-r}, we see, for negative $\beta$, the heat capacity of small black hole is negative while the one of large black hole is positive. We conclude that the small black holes with one event horizon is thermodynamically unstable while the large black holes with one event horizon are stable. For positive $\beta$, the heat capacity of middle black hole is negative while the heat capacity of both  small and large black holes is positive. Thus we conclude that the middle black hole with two horizons is unstable while the small and large black holes with two horizons are stable.

\begin{figure}[htbp]
	\centering
	\includegraphics[width=8cm,height=6cm]{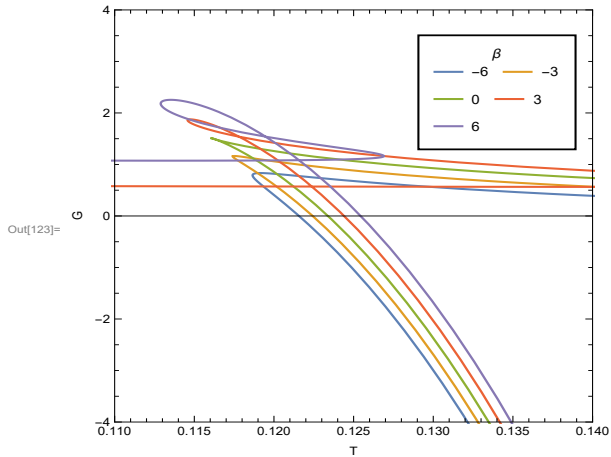}
	\caption{The black hole Gibbs free energy with respect to temperature. The parameters are $n=5$, $\alpha=2$, $a=0.2$, $\lambda=-0.2$.}\label{GT999}
\end{figure}

We can also identify these phase transitions through the diagram of Gibbs energy with respect to black hole temperature. In Fig.~{\ref{GT999}}, we plot the $G(T)$ relations with running $\beta$. It shows that when $\beta<0$, the black holes make phases transition from small black holes to large black holes with the increasing of Hawking temperature. As is known, the specific heat is $C_{Q,P}=-\partial^2G/\partial T^2$. Therefore, the thermodynamically stable and unstable phases have the concave downward and upward $G(T)$ curves, respectively. Then we conclude that the large black holes are thermodynamically stable while the small black holes are unstable. Fig.~{\ref{GT999}} also shows that when $\beta>0$, the system makes phase transitions from small black hole to middle black hole, and finally to large black holes with the increasing of Hawking temperature. In this case, both the phase large and small black holes are thermodynamically stable while the middle black holes are unstable.

\section{Conclusion and discussion}

In conclusion, starting from the n-dimensional black hole solution in Einstein-dilaton gravity and inspired by the four dimensional black hole solution in Einstein-Maxwell-scalar theory, we construct n-dimensional black hole in Einstein-Maxwell-scalar theory. A new coupling function $F(\phi)$ between the scalar field and the Maxwell invariant is present. However, the scalar potential $V(\phi)$ remains the form in Einstein-Maxwell-dilaton gravity. The black hole is described by physical mass, electric charge and two dimensionless coupling constants $\alpha$ and $\beta$.

Then we explore the corresponding thermodynamics. The fundamental thermodynamical functions, namely, the enthalpy, the temperature, the entropy and the thermal volume are derived. Then the Smarr formula is found. It is found that the Smarr formula is related to the dimension of spacetime. But the first law of thermodynamics is not related to dimension of spacetime.

We also study the thermodynamic stability problem and the phase transitions in anti-de Sitter universe. We find, for negative $\beta$, there is generally one event horizon. In this case, there are two phases of black holes, namely, the small black hole phase and large black hole phase, respectively. The small black hole is thermodynamically unstable while the large black hole is stable. With the increasing of Hawking temperature, the system makes phase transitions from small black hole phase to large black hole phase. For positive $\beta$, there are generally two horizons, namely, the inner horizon and outer event horizon. In this case, there are three phases corresponding to small, middle and large black holes, respectively. The middle black hole is thermodynamically unstable while both the small and large black holes are stable. With the increasing of Hawking temperature, the system makes phase transitions from small black hole to middle black hole, and finally to large black holes.

\section*{ACKNOWLEDGMENTS}
This work is partially supported by China Program of International ST Cooperation 2016YFE0100300
, the Strategic Priority Research Program ``Multi-wavelength Gravitational Wave Universe'' of the
CAS, Grant No. XDB23040100, the Joint Research Fund in Astronomy (U1631118), and the NSFC
under grants 11473044, 11633004 and the Project of CAS, QYZDJ-SSW-SLH017.


\bibliographystyle{apsrev4-1}

\bibliography{citationlist}

\end{document}